\documentclass[twocolumn,showpacs,preprintnumbers,amsmath,amssymb,superscriptaddress]{revtex4}
\usepackage{amsmath}
\usepackage{amsfonts}
\usepackage{amssymb}
\usepackage{graphicx}

\begin{document}
\title{The magnetic structure of the $zigzag$ chain family Na$_{x}$Ca$_{1-x}$V$_2$O$_4$ determined by muon-spin rotation}
\author{Oren Ofer}
\email{oren@triumf.ca}
\affiliation{TRIUMF, 4004 Wesbrook Mall, Vancouver, BC V6T2A3, Canada}
\author{Yutaka Ikedo}
 \altaffiliation[Present address: ]
 {Muon Science Laboratory, Institute of Materials Structure Science, KEK,
 1-1 Oho, Tsukuba, Ibaraki 305-0801, Japan}
\affiliation{Toyota Central Research and Development Laboratories Inc., Nagakute, Aichi 480-1192, Japan}
\author{Tatsuo Goko}
\affiliation{TRIUMF, 4004 Wesbrook Mall, Vancouver, BC V6T2A3, Canada}
\author{Martin M\aa{}nsson}
\affiliation{Laboratory for Neutron Scattering, ETH Z\"{u}rich and Paul Scherrer Institute, CH-5232 Villigen PSI, Switzerland}
\author{Jun Sugiyama}
\affiliation{Toyota Central Research and Development Laboratories Inc., Nagakute, Aichi 480-1192, Japan}
\author{Eduardo J. Ansaldo}
\affiliation{TRIUMF, 4004 Wesbrook Mall, Vancouver, BC V6T2A3, Canada}
\author{Jess H. Brewer}
\affiliation{TRIUMF, 4004 Wesbrook Mall, Vancouver, BC V6T2A3, Canada}
\affiliation{Department of Physics and Astronomy, University of British Columbia, Vancouver, Canada V6T2A3}
\author{Kim H. Chow}
\affiliation{Department of Physics, University of Alberta, Edmonton, AB, T6G 2G7 Canada}
\author{Hiroya Sakurai}
\affiliation{National Institute for Materials Science, Namiki, Tsukuba, Ibaraki 305-0044, Japan}

\begin{abstract}
We present muon-spin rotation measurements on polycrystalline samples of 
the complete family of the antiferromagnetic (AF) $zigzag$ chain compounds, 
Na$_x$Ca$_{1-x}$V$_2$O$_4$.  
In this family, we explore the magnetic properties 
from the metallic NaV$_2$O$_4$ to the insulating CaV$_2$O$_4$.  
We find a critical $x_c(\sim0.833)$ which separates the low and high 
Na-concentration dependent transition temperature 
and its magnetic ground state.  
In the $x<x_c$ compounds, 
the magnetic ordered phase is characterized by 
a single homogenous phase and 
the formation of incommensurate spin-density-wave order.  
Whereas in the $x>x_c$ compounds, 
multiple sub-phases appear with temperature and $x$.  
Based on the muon data obtained in zero external magnetic field, 
a careful dipolar field simulation 
was able to reproduce the muon behavior 
and indicates a modulated helical incommensurate spin structure 
of the metallic AF phase.  The incommensurate modulation period obtained by the simulation agrees with that determined by neutron diffraction.
\end{abstract}

\maketitle
\section{Introduction}

Experimental and theoretical studies of 
quasi-one dimensional (quasi-1D) magnets 
have flourished recently.  
On theoretical grounds, such systems seem to experience 
low dimensionality combined with frustrated interactions 
of nearest-neighbors and next-nearest-neighbors competition, 
leading to spin density waves forming chiral order\cite{hikihara} 
and other exotic ground states\cite{zvyagin,oleg}.  
Experimentally, progress has accelerated due to 
recent use of high-pressure techniques 
\cite{varga,akimoto,yamauraChem}, 
enabling synthesis of numerous compounds.  
Such materials show remarkable physical behavior 
\cite{yamauchi, nakamura, mao}, 
where the nature behind these phenomena 
is governed by a strong spin-spin interaction 
along one direction combined with 
a much weaker interaction in the other directions.  
A novel quasi-1D spin system of particular interest 
is NaV$_2$O$_4$ (NVO)\cite{yamaura}, 
which is isostructural with CaV$_2$O$_4$ (CVO)\cite{pieper}.  
Unlike other vanadate compounds, which are spinels\cite{zhang}, 
NVO and CVO posses an orthorhombic structure 
with the {\it Pnma} space group, 
where irregular hexagonal 1D channels are formed 
by a series of edge-sharing VO$_6$ octahedra 
aligned along the $b$-axis.  
The magnetic V ions are thus arranged in a 1D $zigzag$ chain.  
Despite their closely related chemical and crystallographical nature, 
they exhibit dramatically different magnetic and electronic properties.  
The metallic NVO incorporates a mixed valence V$^{+3.5}$ state and 
undergoes an antiferromagnetic (AF) transition at $T_{\rm N}=140$~K, 
with a complex magnetic ground state\cite{yamaura,jun1}.  
In fact, recent neutron studies 
indicate the formation of incommensurate magnetic order 
in NVO below $T_N$\cite{nozaki}.  
The insulating CVO, on the other hand, 
has a V$^{+3}$ valence ($t_{2g}^2, S=1$) 
with an AF transition at $T_{\rm N}=70$~K, 
coexisting with intrachain ferromagnetic interactions.  
Furthermore, the chemical substitution of Na for Ca 
causes an interesting insulator to metal transition 
at a critical Na concentration $x > x_c\sim0.83$ 
in Na$_x$Ca$_{1-x}$V$_2$O$_4$ (NCVO).  
The presence of the insulator to metal transition suggests 
a dynamic change in the electronic structure of NCVO with $x$, 
resulting in the change in its magnetic ground state.  
It is therefore desirable to systematically investigate 
the evolution of the microscopic magnetic order/disorder in NCVO.  

Recently, through the use of 
positive muon-spin rotation and relaxation ($\mu^+$SR), 
we found the formation of static AF order for a few NCVO compounds 
below their $T_{\rm N}$ and proposed a magnetic phase diagram\cite{jun1}.  
$\mu^+$SR is known to be a powerful technique 
for studies on microscopic magnetic nature in solids, 
due to its inherent 100\% spin polarization and local sensitivity.  
The most interesting feature in NCVO is the unusual 
coexistence of AF order and metallic behaviour 
below $T_{\rm N}$ for NCVO with $x > x_{\rm cr}$.  
The main purpose of this paper is, therefore, 
to clarify the microscopic magnetic ground state 
of the entire NCVO family through the use of $\mu^+$SR and 
numerical simulations based on dipolar field calculations.  
This is, to our knowledge, a pioneering attempt 
to deduce the incommensurate modulation period \textit{from} $\mu^+$SR data.  
We also seek to complete the magnetic phase diagram of NCVO 
by additional $\mu^+$SR measurements on insulating and metallic NCVO samples 
with $x = 0.166,~0.41,~0.875$ and $0.958$, 
which were not described in our previous report.  
Our results reveal that the magnetic ground state 
in the insulating compounds with $0 < x\leq x_c$ 
are also an incommensurate magnetically ordered phase.  
This means that not only pure CaV$_2$O$_4$ 
but also the complete family of NCVO 
exhibit incommensurate magnetic order at low $T$,  
concurrent with the change of the valence state 
of the V ion from 3 to 3.5 with increasing $x$.  
This could provide a very interesting challenge for theorists: 
to predict the magnetic phase diagram of this $zigzag$ chain compounds.  

\section{Experiment}

Polycrystalline samples of NCVO were prepared by a solid-state reaction technique under a pressure of 6~GPa using CaV$_2$O$_4$, Na$_4$V$_2$O$_7$, and V$_2$O$_3$ powders as starting materials.
A mixture of the three powders was packed in an Au capsule, then heated at 1300$^{\rm o}$C for 1 hour,
and finally quenched to ambient $T$. A powder X-ray diffraction (XRD) analysis showed that
all the samples were almost single phase with an orthorhombic system, $Pnma$ space group, at ambient $T$. dc-$\chi$ measurements showed that all our samples have almost the same $T$ dependence
as that in the previous report \cite{Sakurai}. The preparation and characterization of the samples
have been reported in greater detail elsewhere \cite{Sakurai}.

The $\mu^+$SR measurements were carried out on the M20 surface muon beamline
with the LAMPF spectrometer at TRIUMF,  the Canada's National Laboratory for Particle and Nuclear Physics located in Vancouver, Canada, which provides a highly intense, polarized beam of positive muons.
The samples, 13 polycrystalline discs with 6~mm diameter and 5~mm thickness of NCVO
, were mounted on the sample holder for the muon veto cryostat insert \cite{sample_holder} by 
Mylar-tape.

Weak transverse field $\mu^+$SR (wTF-$\mu^+$SR)
and zero field $\mu^+$SR (ZF-$\mu^+$SR) spectra were
measured in a non-spin-rotated mode-- i.e.,
the initial muon-spin direction is anti-parallel to the muon momentum direction.
Here,  ``weak" means that the applied field is significantly less than any
 possible spontaneous internal fields ($H_{\rm int}$) in the ordered state.
A wTF-$\mu^+$SR technique is sensitive to local magnetic order
through the $\mu^+$ spin polarization amplitude.
In contrast,  ZF-$\mu^+$SR is a sensitive probe of local magnetic [dis]order
through the precession of the muon due to internal magnetic fields at the muon interstitial sites.

\section{Results}

\begin{figure}[tb]
\includegraphics[width= \columnwidth]{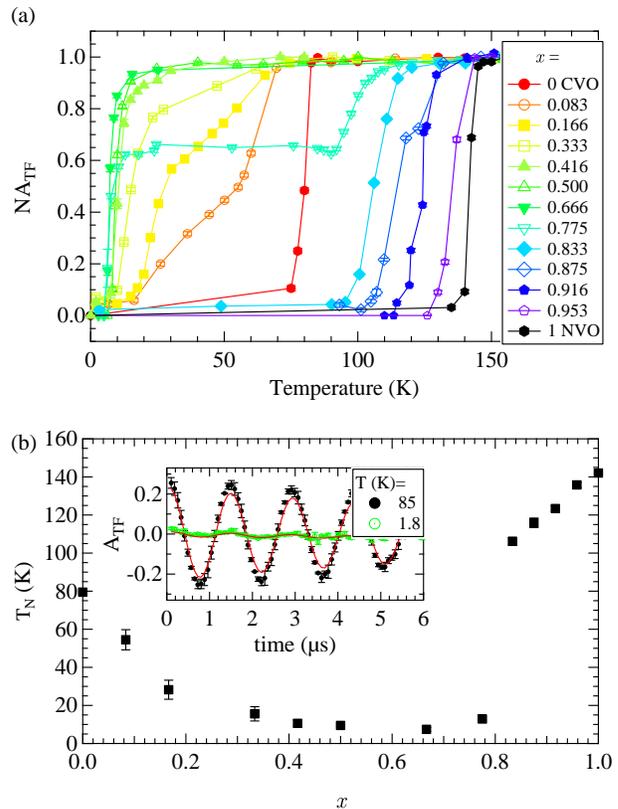}
\caption{(Color online)
(a) Temperature dependence of the normalized
TF asymmetry, $NA_{\rm TF}$, for the complete $x$ family.
(b) The transition temperature, $T_N$ versus the Na concentration, $x$. Inset, a typical wTF-$\mu^+$SR spectrum of the NCVO samples. Shown here the paramagnetic phase, $T=85$, and the ordered state $T=1.8$~K of the $x=0.41$ sample. The solid lines indicates a fit to Eq.~(\ref{tfeq}). }
\label{rawTF}
\end{figure}
In the inset of Fig.~\ref{rawTF}(b) we show the typical wTF asymmetry ($A_{\rm TF}$) represented here by the $x=0.41$ sample at two temperatures,
above and below $T_{\rm N}$.
In the paramagnetic state, at high temperatures
($T>80$~K), the muon-spin precess at the frequency
corresponding to the external transverse field
$H_{\rm TF}=50$~Oe. Below $T_{\rm N}$,
$A_{\rm TF}$ decreases, 
due to the strong static internal magnetic fields, $H_{\rm int}$, obeying $H_{\rm int}>H_{\rm TF}$. The remaining oscillating amplitude, shown in the figure, reflects the portion of muons not coupled to $H_{\rm int}$, thus the fraction of the wTF oscillatory signal below $T_N$ is negligibly small for the magnetic nature of these compounds. The wTF spectrum is well described by an exponentially relaxing cosine oscillation;

\begin{equation}
A_{0}P_{\rm TF}(t)=A_{\rm TF}\exp[-(\lambda_{\rm TF}t)]\cos(\omega_{\rm TF} t+\phi),
\label{tfeq}
\end{equation}
where $\omega_{\rm TF}=\gamma_\mu H_{\rm TF}$
with $\gamma_\mu/(2\pi)=13.554$~kHz/Oe. The fit is represented by the solid lines in the inset of Fig.~\ref{rawTF}(b).

Figure~\ref{rawTF}(a) depicts the temperature dependence of
the normalized $A_{\rm TF}$, $NA_{\rm TF}$,
which is proportional to the fraction of the paramagnetic regions in the sample.
In NCVO, we find two distinct AF phases; one occurs as Na concentration increases up to $x\sim0.8$ and the transition temperature, $T_{\rm N}$ decreases, the second occurs with $x\ge 0.833$, as the Na concentration increases $T_{\rm N}$ takes a sharp turn and increases to higher and higher temperatures with increasing $x$ above $100$~K. 
 These observations are summarized in Fig.~\ref{rawTF}(b) which plots  $T_N$ versus the Na concentration $x$. $T_N$ is found at the middle of the transition and the errors on $T_N$ are taken as $10\%$ of the transition temperature width. Since the $0.083\leq x\leq0.33$ and $x=0.775$ samples do not experience a sharp magnetic transition, we find these samples to be a mixture of phases. 
 Additionally, we can indicate that a critical doping level $x_{\rm cr}\sim0.83$ separates the two phases. 
 However, we wish to emphasize that  both bulk magnetization measurements  and x-ray diffraction analysis
show that each of these samples is almost a single phase with an orthorhombic structure\cite{jun1,varga}.
This clearly indicates the importance of the current microscopic  measurements, which 
show results quite different from the macroscopic bulk measurements.

\subsection{insulating region; NCVO with $0<x\leq0.775$}

\begin{figure}[tb]
\includegraphics[width= \columnwidth]{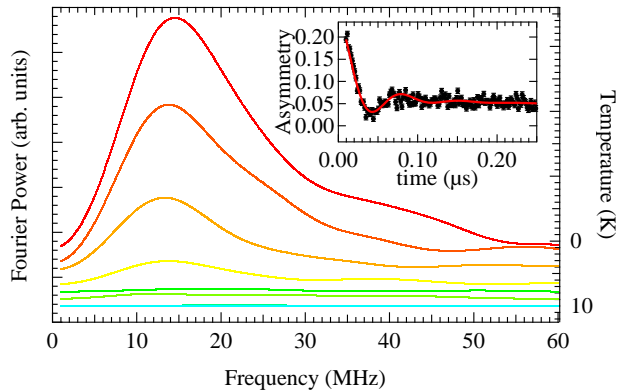}
\caption{(Color online) The apodized ZF-$\mu^+$SR FT of the $x=0.41$ sample. 
Inset shows the time-spectrum the solid line represents a fit to Eq.~(\ref{eq:ZFfit2}).}
\label{lowx}
\end{figure}

We now explore the two AF phases, 
identified by the wTF-$\mu^+$SR measurements,
using the ZF-$\mu^+$SR technique.
The ZF-$\mu^+$SR is a sensitive site-based probe of static magnetism,
in which the time dependent muon polarization signal
is determined exclusively by the $H_{\rm int}$'s
in the sample. 
We expect that different magnetic phases would cause different $H_{\rm int}$s,
resulting in different muon frequencies.
Therefore, following upon the wTF measurements, we performed detailed ZF measurements
for the entire $x$ range. In the $0.41\leq x\leq 0.66$ range,
we concentrate on the low $T$ behaviour, where the AF transition occurs for $T\leq10$~K.
In contrast, for the $x\geq0.833$ samples, we examine a broad $T$ range below 150~K.

Before discussing our current results in NCVO, we briefly 
summarize the ZF-$\mu^+$SR results of pure CVO\cite{jun1}.
The CVO ZF-$\mu^+$SR spectra indicated four temperature dependent frequencies below the AF transition, two of which have significant amplitudes. 
In order to clarify the ground state, dipolar field calculations suggested that AF order 
exists along the two 1D legs in the $zigzag$ chain,
and AF order between the $zigzag$ chains.
This result was found to be consistent with neutron measurements
both for $H_{\rm int}$'s and the magnitude of the ordered V moment.

The typical raw ZF-spectrum, represented by the $x=0.41$ sample, taken at $T=1.8$~K
is depict in the inset of Fig.~\ref{lowx}. One can clearly see a strongly damped oscillation
in the early time domain, $t\leq0.15~\mu$s, at the lowest $T$ measured.
In order to fit the spectrum, we, at first, attempted to apply a combination
of an exponentially relaxing cosine signal and a slowly relaxing signal.
The former corresponds to the static AF ordered signal,
and the latter does to the ``1/3" tail for the powder sample:
\begin{eqnarray}
A_0P_{\rm ZF}(t) &=&
A_{\rm AF}\exp(-\lambda_{\rm AF} t)\cos(\omega_{\rm AF} t+\phi_{\rm AF})
\cr
&+& A_{\rm tail}\exp(-\lambda_{\rm tail} t).
\label{eq:zfeq1}
\end{eqnarray}
\begin{figure}[ptb]
\includegraphics[width= \columnwidth]{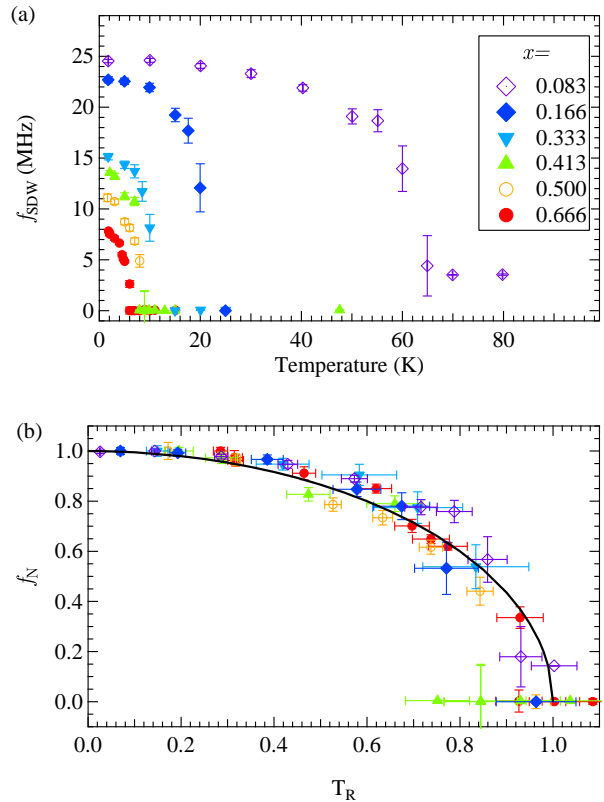}
\caption{(Color online) (a)
The ZF SDW frequency [$f_{\rm SDW}\equiv\omega_{\rm SDW}/(2\pi)$]  versus temperature and
(b) the normalized $f_{\rm SDW}$, $f_N$, versus the reduced temperature [see text]. The solid line indicates the $T$ dependence of the BCS gap energy. }
\label{zfsdw}
\end{figure}

Although the ZF-spectrum was reasonably fitted by Eq.~(\ref{eq:zfeq1}),
the initial phase ($\phi_{\rm AF}$) delays by
$(66\pm2)^{\circ}$, $(57\pm9)^{\circ}$ and
$(44\pm6)^{\circ}$ in the $x=0.41$, $0.5$ and $0.66$ samples.
Since $\phi_{\rm AF}$ should be zero for a simple commensurate AF order,
the delay indicates either a wide distribution of $H_{\rm int}$ or
a formation of an incommensurate (IC) AF order.
In fact, the Fourier Transform (FT) of the ZF time-spectrum
in $0.41\le x\le0.66$ clearly demonstrates
the wide field distribution at the muon sites [see
the main panel of Fig.~\ref{lowx} as an example of the $x=0.41$ sample].
The FT spectrum also indicates
the presence of a shoulder on the higher frequency side,
suggesting the formation of complex magnetic order.
Indeed, the ZF-spectrum is better fitted by a zeroth-order Bessel function of the first kind
[$J_0(\omega_{\mu}t$)],
especially in explaining the fast relaxing behaviour in the early time domain
(before 0.02~$\mu$s).
A ZF-spectrum described by $J_0$ is well established signature that
the $\mu^+$'s experience an IC magnetic field in the lattice \cite{muSR_ICfields}.
Hence, the ZF-$\mu^+$SR spectra at low $T$ were fitted with a combination of two signals;
\begin{eqnarray}
A_{0} P_{\rm ZF}(t) &=&
  A_{\rm SDW} \exp(-\lambda_{\rm SDW} t) J_0(\omega_{\rm SDW}t)
 \cr
 &+& A_{\rm tail}\exp(-\lambda_{\rm tail} t).
   \label{eq:ZFfit2}
\end{eqnarray}
The fit is demonstrated by the solid line in the inset of Fig.~\ref{lowx}.
In Fig.~\ref{zfsdw}(a), we plot the frequency $f_{\rm SDW}\equiv\omega_{\rm SDW}/2\pi$, extracted from the fits, versus $T$.
For all $x\le0.66$, as $T$ increases, $f_{\rm SDW}$ decreases with 
an increasing slope
(d$f_{\rm SDW}$/d$T$) and approaches zero at $T_{\rm N}$.
Interestingly, as indicated in Fig. 3b, the normalized $f_{\rm SDW}$
[$f_{\rm N}=f_{\rm SDW}(T)/
f_{\rm SDW}(T\rightarrow 0~{\rm K})$]
and the reduced $T$ ($T_{\rm r}\equiv T/T_{\rm N}$)
are independent of $x$ [see Fig.\ref{zfsdw}(b)].
In fact, the $\omega_{\rm N}$-vs.-$T_{\rm r}$ curves
now collapse to a single universal curve,
indicating the same origin of the IC transition.
This universal curve is well explained by the $T$ dependence of the {\sf BCS} gap energy,
as expected for the order parameter of the IC-AF state \cite{Gruner_1}\cite{Le}. This was recently found by a $^{51}$V NMR study, which suggested an IC AFM for $x\geq 0.77$\cite{hikaru} (Note that AF resonance lines in the metallic phase were observed in Ref.\cite{hikaru}).  We therefore find that the IC phase in the insulating region includes all compounds, at low $T$.   

The transverse relaxation rate, $\lambda_{\rm SDW}$, shows typical critical
behavior as $T \to T_N$, namely, $\lambda_{\rm SDW}$ increases with increasing
$T$ up to $T_N$ as expected \cite{ryan,uemura,kalvius}.  The longitudinal
relaxation rate, $\lambda_{\rm tail}$, can in principle reveal whether
frustration prevents the system from reaching the full static limit, as observed
in some highly frustrated systems: normally one expects $\lambda_{\rm tail} \to 0$ (static limit)
as $T \to 0$.  However, $\lambda_{\rm tail}$ is very small ($<
1\mu$s$^{-1}$) compared with $\lambda_{\rm SDW}$ and difficult to
determine reliably, especially in the presence of several signals, hence we cannot
make a confident statement.

\subsection{metallic region; NCVO with $x\ge0.83$}

\begin{figure}[ptb]
\includegraphics[width= \columnwidth]{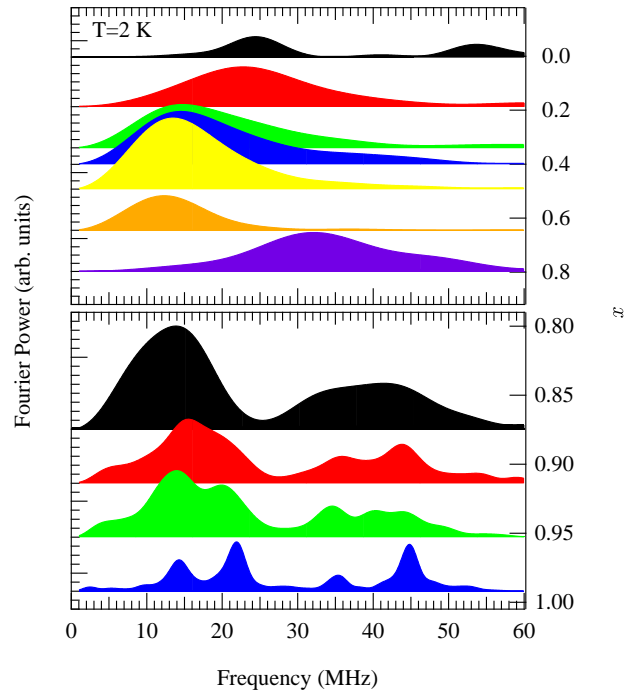}
\caption{(Color online)The apodized ZF FT of the whole $x$ range, 
top displays the insulating region characterized by a single wide frequency distribution, 
bottom shows the metallic region characterized by numerous frequencies. }
\label{raw10to12a}
\end{figure}
\begin{figure}[ptb]
\includegraphics[width= \columnwidth]{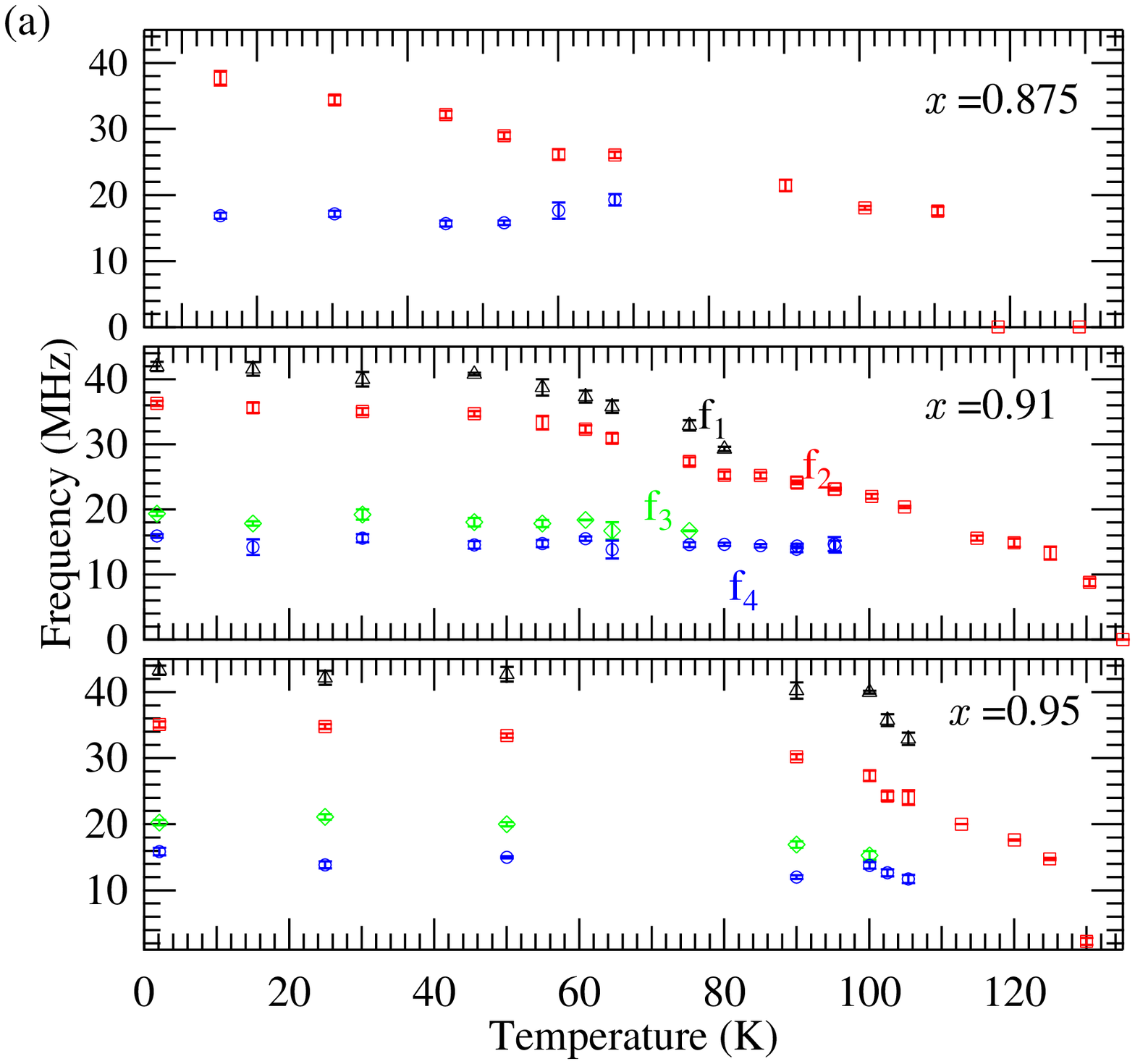}
\includegraphics[width= \columnwidth]{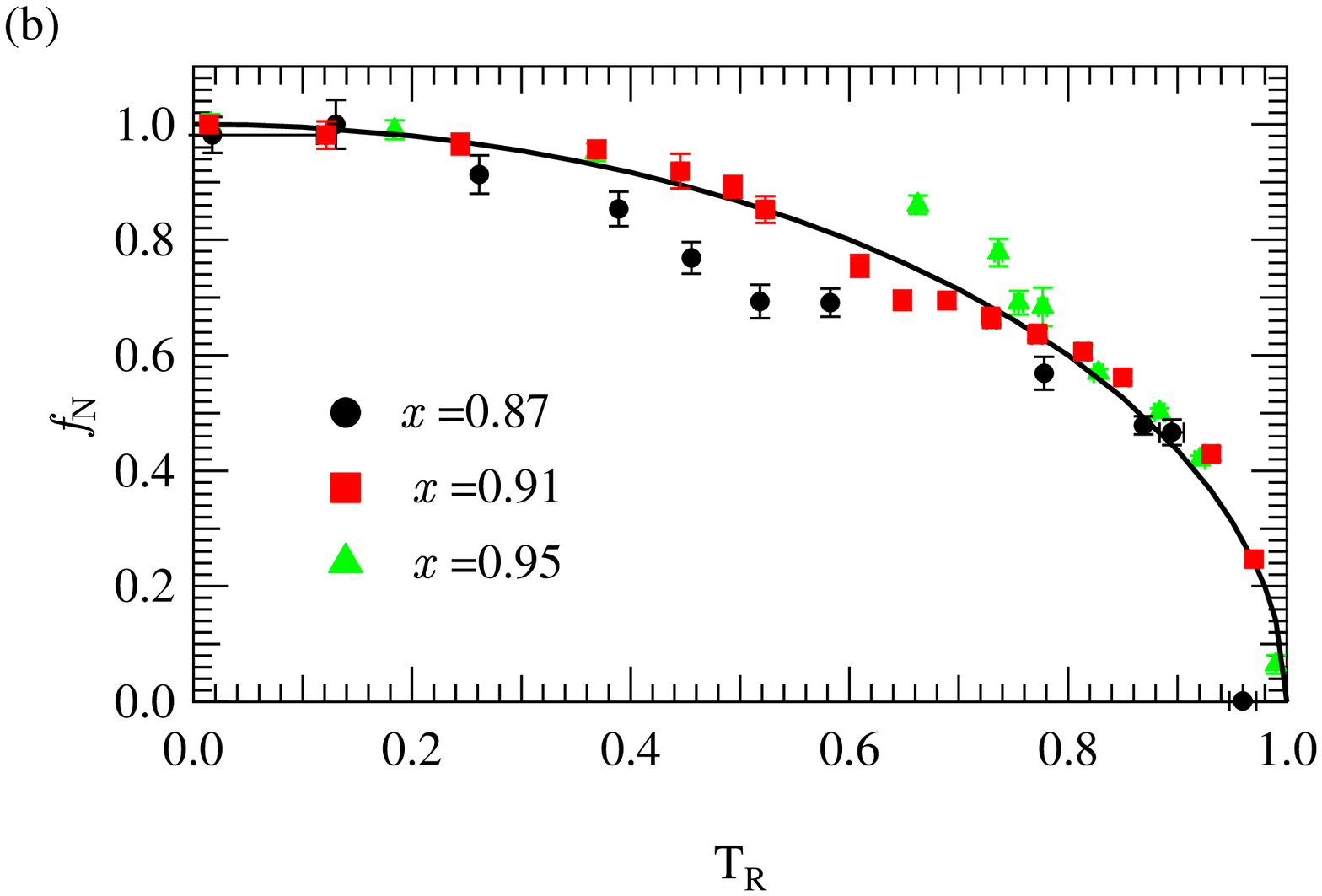}
\caption{(Color online) (a) The temperature dependence of the frequencies in the metallic phase, top to bottom,  $x=0.875,~0.91,~0.95$. (b) The  the reduced-temperature ($T_r$) dependence of the normalized $f_2$ frequency, $f_\text{N}$. The solid line demonstrates the BCS gap enegy.}
\label{raw10to12b}
\end{figure}

\begin{figure}[tb]
\includegraphics[width= \columnwidth]{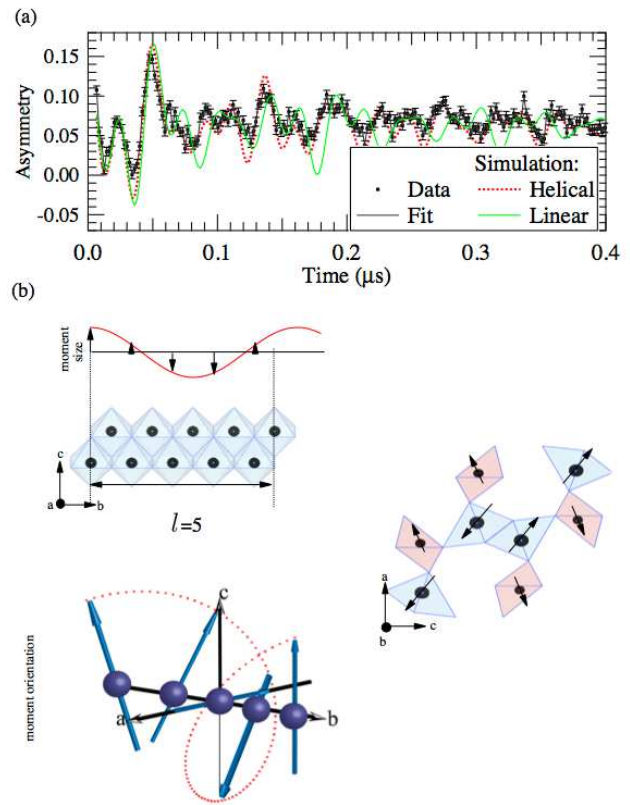}
\caption{(Color online) (a) The raw ZF-$\mu^+$SR data of NaV$_2$O$_4$.
The solid (grey) line indicate a fit, the helical simulation is represented by the dotted line (red), and the linear by the dashed (green) line. (b) The suggested magnetic structure, the moment orientation in the $\hat{ac}$ plane (right), the moment size and the helical modulation length $l$ (center), the magnetic $V$ ions and their moments in $l=1$. } \label{simulation}
\end{figure}

The FT of the whole $x$ range, at $T=2$~K, is presented in Fig.~\ref{raw10to12a}. Clearly, the ZF measurements of the metallic compounds, i.e., $x\ge0.83$ samples, reveal different behavior
than those at smaller x.
Unlike the insulating compounds, the metallic compounds, display multiple frequencies (two in $x=0.875$, four in $x\ge0.9$), forming one frequency at higher $T$  and vanishing above the phase transition (not shown).
Therefore, the ZF-spectra for these compounds were fitted by a sum of four oscillating signals with a slow exponential relaxation, 
\begin{eqnarray}
A_0P_{\rm ZF}(t) &=& \sum_{i=1}^nA_i\cos(\omega_i t+\phi_i)\exp(-\lambda_i t)
\cr
&+& A_{\rm tail}\exp(-\lambda_{\rm tail} t)
\label{eq:zfeq3}
\end{eqnarray}
where $n=4$ and $\omega_i, \phi_i$, and $\lambda_i$ are
the angular frequency, the initial phase, and the relaxation rate
of the $i$-th muon precession, $A_{\rm tail}$ and $\lambda_{\rm tail}$ are the asymmetry and the relaxation rate for the ``1/3 tail" signal\cite{ishida,zorkoJAP}, as we measured powder samples.

The quality of the fit, shown for NVO,
is represented by the solid green line in Fig.~\ref{simulation}(a).
The $T$ dependence of the frequencies, $f_i\equiv\omega_i/(2\pi)$, of the $0.83\leq x\leq 0.95$ samples 
shown in Fig.~\ref{raw10to12b}(a). In the $x=0.83$ sample,
we note that at $T<70$~K two frequencies are distinguished, but
at $70\le T\le110$~K a single frequency is observed. In $x=0.91$ ($x=0.95$),
the number of frequencies is decreased from 4, at $T<70$K
to 2 (3) and then to one frequency.
Recent neutron scattering measurements suggests that
a small displacement of O$^{2-}$ ions, which leads to a change in the muon sites\cite{nozaki} is the trigger to this intermediate phase. This scenario was also proposed in other compounds\cite{jun2007}.
One of the distinctive features in this $x$ range,  are two common frequencies. At base $T$,
$f_2\sim 35$MHz and $f_4\sim17$MHz are $x$ independent.  As in the insulating phase, by plotting the frequency versus the reduced temperature results in the collapse of these frequencies to a single curve, suggesting the IC AFM in this phase as well. Fig.~\ref{raw10to12b}(b) demonstrates this for $f_2$. In the next section we aim to find the correct spin configuration which produces this behavior.
\begin{table}
\begin{tabular}{p{.2\columnwidth}p{.2\columnwidth}p{.2\columnwidth}p{.2\columnwidth}}
\hline\hline
& $x$ & $k$[$\hat b$] (\AA$^{-1}$) & $\chi^2$  \\ \hline
Helical &0.916 & 0.381 & 1.591 \\
&0.958 & 0.363 & 3.732\\
&1 & 0.191 & 0.910\\
\hline
Linear &1 &  0.571 & 105.203\\
\hline\hline
\end{tabular}
\caption{The helical simulation result showing the modulation length along $\hat{b}$, $l$, and the goodness-of-fit, $\chi^2$ for Na-concentrations $x$.} \label{simtable}
\end{table}

\section{Discussion}
\subsection{Incommensurate AF structure for NVO}

We now address the origin of the 4 frequencies observed in metallic phase
using computer simulations. We aim to fit the raw experimental data with a minimal number of free parameters. We use the known crystallographic structure in order to create the simulated crystal model\cite{Sakurai}. We then find all possible muon sites in the sample using electrostatic potential calculations\cite{explain}. Indeed the model  finds 4 distinct muon sites, which was also seen previously in the pure NVO\cite{jun1}. The next step was to find the correct spin configuration which results in the fields which give the muon precession measured in each site. Hence, dipolar fields were calculated with different independent alignments of the AF moments of the two $zigzag$ chains, residing in the $\hat{ac}$ plane.
The simulation also probed different IC helical modulations,
that is the magnetic moment size at the $l^{th}$ unit cell is
\begin{equation}
{\bm m}={\bm m}_0\cos(2\pi{\bm k}\cdot{\bm l})
\end{equation}
where ${\bm k}$ is the propagation vector along $\hat{b}$
and ${\bm m}_0$ is the V moment size, which is the only free parameter, and is given in Ref.\cite{jun1,nozaki}. We average over degenerate states obtained by the simulation in order to accurately describe this model. We use $\chi^2$ criteria as a crude guide in selecting the potential realized state. For comparison, we also performed the same calculation
for a linear IC spin density wave, which is proposed by neutron measurements\cite{nozaki}.
However, as seen in Table I, $\chi^2$ for a linear IC-SDW model was always higher than a helical-modulated setup. Figure~\ref{simulation}(a) displays the experimental ZF-spectrum of NVO
taken at $T=2$~K with the simulated spectrum assuming both the helical and linear configuration.  

We find that the best fit to the experimental data is,
thus, a configuration with an helical IC modulation along $\hat{b}$ axis. Table \ref{simtable} summarizes the simulation results and shows the propagation vector $k$ versus $x$ and the $\chi^2$ goodness-of-fit to the $T=2$~K experimental data of these compounds.
The value of NVO is found to be very consistent with that obtained
by the recent neutron experiment\cite{nozaki},
namely, the modulation period is $5.214\times b$ along the $b$-axis.
However, the powder neutron diffraction result does not suggests a helical IC order
but rather a linear IC-SDW order in NVO. The reason for this discrepancy is not clear. We speculate that the since the neutron result was obtained from powder samples through the Rietveld analyses, and the result published may be that of a local minima. Secondly, it might be that the dipolar field alone, which do not take into account crystal fields might not be fully appropriate to describe such a subtle magnetic state.
In fact, a helical IC ordered state is more reasonable in
explaining the coexistence of the AF order and metallic conductivity below $T_N$ for NVO. 
Hence, $\mu^+$SR is found to play a significant role to determine the AF structure
in NCVO. In Fig.~\ref{simulation}(b), we plot the corresponding magnetic structure suggested by the $\mu^+$SR simulation.

\begin{figure}[tb]
\includegraphics[width=\columnwidth]{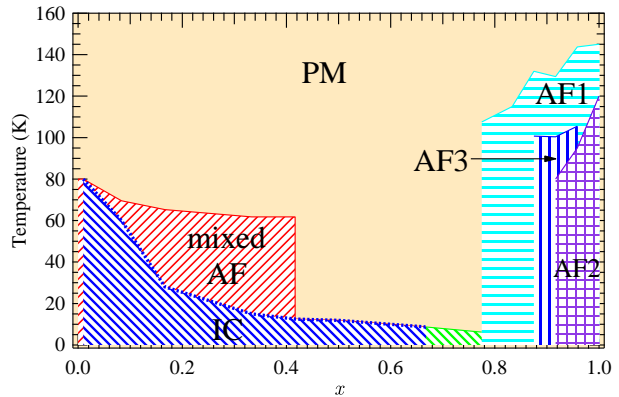}
\caption{(Color online) The phase diagram of Na$_x$Ca$_{1-x}$V$_2$O$_4$ family. PM: Paramagnetic, AF: Antiferromagnetic, IC: Incommensurate.}
\label{phasediag}
\end{figure}

\subsection{Phase diagram}

To summarize the muSR measurements of NCVO carried out in
the present work, 
we propose in Fig.~\ref{phasediag} the complete phase diagram as a function of x.
In particular, the IC-AF phase with its wide field distribution, is present
at low temperature of $0<x\leq0.66$.
Interestingly, (above the IC phase), the compounds
$0<x\leq0.41$ show a mixture of CVO and the $x=0.416$ compound by the TF measurements (shown as mixed-AF in Fig. \ref{phasediag}),
but such mixture behavior is absent in the ZF data, in which only the IC phase is identified. 
The metallic phase, $x\ge0.83$, is characterized by numerous AF phases. AF1 indicates the AF order identified by a single frequency, found below $T_N$ for $x=0.83$ and at higher temperatures for $x\ge0.875$. AF2 and AF3 describes the low temperature AF order characterized by 4 frequencies occurring in the $x\ge 0.91$. We therefore updated the previously published phase diagram\cite{jun1} with the introduction of $x=0.41$, identified the IC phase at $x\le0.66$, as well as the new data on $x=0.875,~0.958$ shed additional light on the multiple AF phases.  

\section{Summary}
In summary, $\mu$SR measurements on the Na$_{x}$Ca$_{1-x}$V$_2$O$_4$ family reveal
a diverse number of magnetic phases for different $x$.
Compounds in the insulating phase, $0<x<0.8$, indicate a single muon site,
 data analyses identifies an incommensurate spin density wave phase.
 Whereas in the metallic phase, at high Na concentrations, $x>0.8$, data suggests several muon sites. Numerical simulations verifies 4 magnetically distinct muon sites, and validates the existence of an incommensurate helical spin density of the V $zigzag$ chain,
with a spin-wave propagation vector $\bf{k}$ along $\hat{b}$ axis.
Finally, we wish to point out that,  although $\mu^+$SR is usually thought to be incapable
to determine the correlation length of the magnetic order, the present study clearly 
demonstrates that $\mu^+$SR  plays a crucial role for clarifying the magnetic structure of NCVO.

\begin{acknowledgments}
We thank the staff of TRIUMF for help with the $\mu^+$SR experiments.
YI and JS are partially supported
by the KEK-MSL Inter-University Program for Oversea Muon Facilities, and
JHB is supported at UBC by NSERC of Canada,
and (through TRIUMF) by NRC of Canada,
KHC by NSERC of Canada and (through TRIUMF) by NRC of Canada, and
HS by WPI Initiative on Materials, Nanoarchitronics, MEXT, Japan.
This work is also supported by Grant-in-Aid for Scientific Research (B),
19340107, MEXT, Japan.
\end{acknowledgments}


\begin{thebibliography}{99}
\bibitem{hikihara} Toshiya Hikihara, Lars Kecke, Tsutomu Momoi and Akira Furusaki,  Phys. Rev. B \textbf{78}, 144404 (2008). 
\bibitem{zvyagin} A. A. Zvyagin and S.-L. Drechsler, Phys. Rev. B \textbf{78}, 014429 (2008)
\bibitem{oleg} O. Tchernyshyov, Phys. Rev. Lett. \textbf{93}, 157206 (2004).
\bibitem{varga} Tamas Varga, John F. Mitchell, Kazunari Yamaura, 1, David G. Mandrus and Jun Wang, Solid State Sciences \textbf{11}, 694 (2009). 
\bibitem{akimoto} Junji Akimoto, Junji Awaka, Norihito Kijima, Yasuhiko Takahashi, Yuichi Maruta, Kazuyasu Tokiwa, Tsuneo Watanabe, J. Sol St. Chem., \textbf{179}, 169 (2006). 
\bibitem{yamauraChem} Kazunari Yamaura,  Qingzhen Huang,  Lianqi Zhang,  Kazunori Takada,  Yuji Baba, Takuro Nagai, Yoshio Matsui, Kosuke Kosuda, and Eiji Takayama-Muromachi, J. Am. Chem. Soc. \textbf{128}, 9448 (2006).
\bibitem{yamauchi} T. Yamauchi, Y. Ueda and N. Mori, Phys. Rev. Lett. \textbf{89}, 057002 (2002).
\bibitem{nakamura} M. Nakamura, A. Sekiyama, H. Namatame, A. Fujimori,  H. Yoshihara, T. Ohtani, A. Misu, M. Takano, PRB \textbf{49}, 16191 (1994).
\bibitem{mao}  Z. Q. Mao, T. He, M. M. Rosario, K. D. Nelson, D. Okuno, B. Ueland, I. G. Deac, P. Schiffer, Y. Liu, and R. J. Cava, PRL 90, 186601 (2003).
?\bibitem{yamaura} K. Yamaura, M. Arai, A. Sato, A. B. Karki, D. P. Young, R. Movshovich, S. Okamoto, D. Mandrus, and E. Takayama-Muromachi, Phys. Rev. Lett. \textbf{99}, 196601 (2007). 
\bibitem{pieper}  O. Pieper, B. Lake, A. Daoud-Aladine, M. Reehuis, K. Proke$\check{\rm s}$, B. Klemke, K. Kiefer, J. Q. Yan, A. Niazi, D. C. Johnston, and A. Honecker, Phys. Rev. B \textbf{79}, 180409(R) (2009). 
\bibitem{zhang} Z. Zhang, Despina Louca, A. Visinoiu, and S.-H. Lee, J. D. Thompson, T. Proffen, and A. Llobet, Y. Qiu, S. Park and Y. Ueda,  Phys. Rev. B 74, 014108 (2006). 
\bibitem{jun1} Jun Sugiyama, Yutaka Ikedo, Tatsuo Goko, Eduardo J. Ansaldo, Jess H. Brewer, Peter L. 
Russo, Kim H. Chow, and Hiroya Sakurai, Phys. Rev. B \textbf{78}, 224406 (2008). 
\bibitem{nozaki} Hiroshi Nozaki, Jun Sugiyama, Martin M\aa nsson, Masashi Harada, Vladimir Pomjakushin, Vadim Sikolenko, Antonio Cervellino, Bertrand Roessli, and Hiroya Sakurai, Phys. Rev. B \textbf{81}, 100410(R) (2010).
\bibitem{Sakurai} Hiroya Sakurai, Phys. Rev. B \textbf{78}, 094410 (2008).
\bibitem{sample_holder}http://musr.ca/equip/hold/mvhold.html
\bibitem{muSR_ICfields}  J. Major, J. Mundy, M. Schmolz, A. Seeger, 
K. -P. D$\ddot{\rm o}$ring, K. F$\ddot{\rm u}$rderer, M. Gladisch, D. Herlach  and G. Majer, Hyperfine Interactions \textbf{31}, 259 (1986); A. Amato, R. Feyerherm, F. N. Gygax, A. Schenck, H. v. L$\ddot{\rm o}$neysen and H. G. Schlager, Phys. Rev. B \textbf{52}, 54 (1995); N. Papinutto, M.J. Graf, P. Carretta, A. Rigamonti, M. Giovannini, Physica B 359, 89 (2005).
\bibitem{Gruner_1} G. Gr\"uner, Density Waves in Solids, Addison-Wesley-Longmans, Reading, 1994 (Chapter 4), and references cited therein.
\bibitem{Le}L. P. Le, G. M. Luke, B. J. Sternlieb, W. D. Wu, Y. J. Uemura, J. H. Brewer, T. M. Riseman, R. V. Upasani, L. Y. Chiang and P. M. Chaikin, Europhys. Lett. \textbf{15}, 547 (1991).
\bibitem{hikaru} Hikaru Takeda, Masayuki Itoh, and Hiroya Sakurai, J. Phys.: Conf. Ser. \textbf{200}, 012200 (2010).
\bibitem{ryan} D. H. Ryan, J. M. Cadogan, J. van Lierop, Phys. Rev. B \textbf{61}, 6816 (2000).
\bibitem{uemura} Y. J. Uemura, T. Yamazaki, D. R. Harshman, M. Senba and E. J. Ansaldo, Phys. Rev. B \textbf{31}, 546 (1985).
\bibitem{kalvius} G. M. Kalvius, D. R. Noakes, and O. Hartmann, 
{\it Handbook on the Physics and Chemistry of Rare Earths}  
edited by K. A. Gschneidner Jr., L. Eyring, and G. H. Lander, 
(North-Holland, Amsterdam, 2001) vol. 32, chap. 206.
\bibitem{ishida} T. Ishida, S. Ohira, T. Ise, K. Nakayama, I. Watanabe, T. Nogami and K. Nagamine, Chem. Phys. Lett. \textbf{330}, 110 (2000). 
\bibitem{zorkoJAP} A. Zorko, M. Pregelj, H. Berger and D. Arcon, J. of Appl. Phys. \textbf{107}, 09D906 (2010).
\bibitem{jun2007}  Jun Sugiyama, Yutaka Ikedo, Peter L. Russo, Hiroshi Nozaki, Kazuhiko Mukai, Daniel Andreica, Alex Amato, Maxime Blangero, and Claude Delmas, Phys. Rev. B \textbf{76}, 104412 (2007). 
\bibitem{explain} The unit cell was electostatically mapped using the formal charge states for each of the ions in the unit cell. The result indicates 4 inequivalent magnetic sites next to O$^{-2}$ suggesting a formation of a $\mu$-O covalent bond.
\end{thebibliography}
\end{document}